\documentclass[12pt]{article}
\usepackage{graphicx}
\begin{document}

\begin{center}

{\bf \Large Faint Companions of Isolated 2MIG Galaxies}

\bigskip
                                                                                  
{\bf V. E. Karachentseva$^1$ , I. D. Karachentsev$^2$ , and
O. V. Melnyk$^{3, 4}$}\\
\end{center}

$^1$Main Astronomical Observatory, National Academy of Sciences, ul.
Acad. Zabolotny 27, Kiev, 03680 Ukraine \\
$^2$Special Astrophysical Observatory, Russian Academy of Sciences,
Nizhnii Arkhyz, 369167 Russia
$^3$ Institut d'Astrophysique et de Geophysique, Universit\'{e} de Li\`{e}ge,
All\'{e}e du 6 Ao\^{u}t, 17, B5C B4000 Li\`{e}ge, Belgium\\
$^4$ Astronomical Observatory, Taras Shevchenko University, Observatornaya
3, Kiev, 04053 Ukraine\\

\bigskip

{\bf \large Abstract}

\bigskip
We present the results of a search for companions around the isolated
galaxies from the 2MIG catalog. Among 3227 2MIG galaxies we detected 125
objects with a total of 214 neighbors having radial velocity differences
of $\Delta V<500$ km/s and projected separations of $R_p<500$ kpc relative
to the 2MIG galaxies. The median luminosity of the companions is 1/25 of
the luminosity of catalog galaxies, which has little effect on the dynamic
isolation of the latter. The median ratio of the orbital mass to the
K-luminosity determined from 60 companions of E and S0 2MIG galaxies,
$63M_{\odot}/L_{\odot}$, is signficantly greater than that found from
the spiral galaxy companions ($17M_{\odot}/L_{\odot}$). We note that a
fraction of 2MIG galaxies with companions may be a part of low-contrast
diffuse structures: clouds and filaments.
\bigskip


\bigskip

{\bf \large 1. INTRODUCTION}

\bigskip

 Measurements of radial velocities and projected separations of dwarf
galaxies, located in the vicinity of normal isolated galaxies have been
used by many authors to determine the masses of central galaxies and dark
matter density profile around them. Zaritsky et al. [1] and Herbert-Fort
[2] measured radial velocities and distances of 69 and 78 companions around
 the isolated Sb-Sc spirals, respectively, and estimated the typical masses
 of dark haloes up to the galactocentric distance of about ~250 kpc. The
capabilities of this approach have considerably expanded with the publication
 of the Sloan Sky Survey, SDSS [3] and other spectral surveys of broad areas
of the sky: 2dFGRS [4] and 2MRS [5]. The kinematic features of the companions
 depending on the luminosity and morphology of central galaxies, as well as
their redshifts z were studied in [6--10]. These authors used different ways
of identifying both the isolated galaxies and their companions. According to
[6], the dispersion of radial velocities of the companions decreases
approximately from 120 km/s at the projected separation of about 20 kpc to
around 60 km/s at the distance of approximately 350 kpc, and increases with
luminosity of the central galaxy as $\sigma_v\propto L^{0.3}$ , while the
orbital mass-to-luminosity ratio increases as $M_{orb}/L\propto L^{0.5}$.
Norberg et al. [9] noted that the dispersion of radial velocities of the
companions does not essentially depend on the projected separation, and
the orbital mass-to-blue (B-band) luminosity ratio is about
$\sim36M_{\odot}/L_{\odot}$ and $\sim180M_{\odot}/L_{\odot}$ in spiral and
elliptical galaxies, respectively. Investigating the surface density of a
number of companions in the SDSS survey, Chen et al. [7] demonstrated that
it $1.7\pm0.1$ decreases with projected separation $R_p$ as $R_p$
independently of the luminosity of central galaxy or the luminosity of
companions. Similarly, the observational data on the companions were used
to study the characteristics of ``fossil'' groups, where the central galaxy
has an order higher luminosity than that of its faint companions [11--13].

It should be noted that there is a significant discrepancy between the
results and conclusions of different authors. Its nature mainly lays in the
requirements applied to the degree of isolation of studied galaxies, which
affects the relative number of optical (background) ``companions'' and,
henceforth, the linear dimensions of the subsystem of physical companions
and the radial velocity dispersion in it.

Compiling the 2MIG catalog of
isolated galaxies [14], we paid attention to the cases where a well-isolated
galaxy had small companions with radial velocities close to it. In this
paper, we set ourselves to organize and analyze the data on such systems.

\bigskip

{\bf \large 2. ISOLATED GALAXIES OF THE 2MIG CATALOG}

\bigskip

 The first catalog of isolated galaxies of the northern sky (KIG) [15]
included 1050 galaxies brighter than $m_{ph}=15.5^m$, around which the
neighboring ``significant'' galaxies with angular diameters of $a_i$ in
the range of $a_1/4<a_i<4a_1$  satisfied the condition $x_i/a_i>20$. Here
$a_1$ means the standard angular diameter of the potentially isolated galaxy
and $x_i$ is the angular (projected) distance of its neighbor ``i''. The
research has shown [16] that this simple criterion proved to be quite
effective for the selection of dynamically isolated galaxies. With the
advent of a photometrically homogeneous IR survey of the entire sky, the
Two Micron All-Sky Survey (2MASS), [17], and a catalog of extended IR
sources, 2MASS XSC [18], a similar approach was used to compile the 2MIG
(2MASS Isolated Galaxies) [14] Catalog. The catalog includes galaxies with
apparent $K_s$ -magnitudes in the range of $4.0\leq K_s\leq 12.0^m$  and
infrared angular diameters $a_K\geq30^{\prime\prime}$, in cases when all
their significant neighbors with angular diameters $a_i$
 and angular distances $x_i$ satisfy the condition $x_i/a_i>30$. A more
stringent constraint on the dimensionless distances   $x_i/a_i$ is
conditioned by the fact that the IR diameter of a galaxy is on the average
one and a half times smaller than its standard diameter. Since the 2MASS
survey is not very sensitive to the blue galaxies of low surface brightness,
during the selection of isolated galaxies we made an additional visual
inspection of the neighborhood in the optical DSS survey images, and used
the original criterion [15] to the neighboring ``significant'' galaxies.
We also checked the radial velocities of significant neighbors, visible in
the DSS survey, excluding the galaxy candidates with radial velocities close
to their neighbor's velocity ( $\mid\Delta V\mid<500$ km/s).

As a result of
such a multistage selection process, we compiled the 2MIG catalog of isolated
 galaxies, which includes 3227 objects. The effective depth of the catalog,
about ~6500 km/s, is about the same as that of the KIG. The main population
of the 2MIG catalog is composed of spiral galaxies (80\%), about 19\%
accounts for E and S0 objects, while the fraction of irregular and BCD
galaxies does not exceed 1\%.

Various optical and hydrogen characteristics
of the sample of 2MIG galaxies were considered in [19]. According to the
authors [19], the completeness of the catalog is about 80\% up to the
limiting apparent
magnitude of $K_s = 11.5$. The isolation criterion selects into the 2MIG
catalog 6.2\% of the total number of 51572 galaxies with apparent magnitudes
of $K_s < 12.0^m$ , and angular diameters of $a_K>30^{\prime\prime}$.

Currently, more than 70\% of 2MIG galaxies have measured radial velocities.
The comments to the catalog note the cases ($N \sim140$ when near a given
isolated galaxy, fulfilling the isolation criterion, there were insignificant
 neighbors with radial velocity difference of  $\mid\Delta V\mid<500$ km/s
with respect to the given galaxy. Such cases were examined more thoroughly,
and the results of these examinations are presented below.

\bigskip

{\bf \large 3. A SEARCH FOR FAINT COMPANIONS AROUND 2MIG\\ GALAXIES}

\bigskip

 The ongoing massive sky surveys and radial velocity measurements in ever
more fainter galaxies lead to the discovery of dwarf companions of galaxies,
which seemed to be absolutely isolated. In most cases, the galaxies remain
to be dynamically isolated, since the presence of new small physical
companions does not violate the adopted isolation conditions. An example of
this situation is a well-isolated BCD galaxy NGC 1156 = KIG 121 = 2MIG 360,
having apparent magnitudes of $B = 12.32^m , K_s = 9.54^m$ and the
heliocentric radial velocity of $V_h = 376$ km/s. Within the HI survey AGES,
carried out with the Arecibo radiotelescope, a dwarf galaxy AGES
J030039+254656 was discovered [20] with an apparent magnitude of $B = 18.1^m$
 and radial velocity of $V_h = 308$ km/s, located at an angular distance of
$35^{\prime}$ (80 kpc) from NGC 1156. There is every reason to call this
dwarf galaxy a physical companion of NGC 1156, the presence of which, however,
 does not distort the dynamic autonomy of this bright galaxy.

The emergence
of new data on radial velocities of galaxies allows making an a posteriori
estimate of the eficiency of the isolation criterion used. Viewing the
vicinities of 2MIG galaxies, we selected among their neighbors the galaxies
with radial velocity difference of $\mid\Delta V\mid<500$ km/s and projected
separation of $R_p < 500$ kpc relative to the principal (2MIG) galaxy without
restrictions on magnitude difference between the main galaxy and its
companion.

As a result of our examination using the NED database
(http://nedwww.ipac.caltech.edu), we identificated 125 2MIG galaxies with
companions, the total number of which amounted to N = 214. The data on 339
these galaxies are presented in the Table. Its columns contain: (1) the number
of a given galaxy in the 2MIG catalog and the presence of its companion/neighbors
(in the following lines); (2) equatorial coordinates of galaxies at the epoch

\begin{figure}
\includegraphics{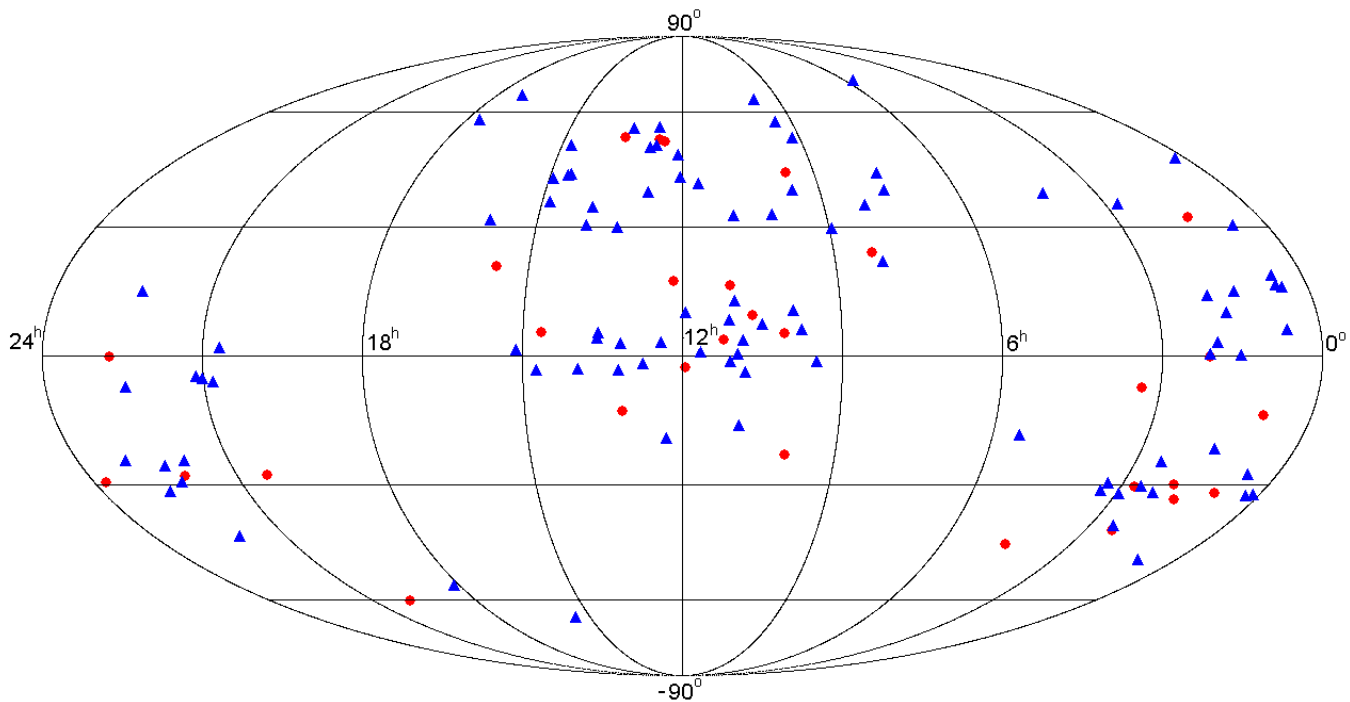}
\caption{The distribution of 125 2MIG galaxies with companions in the sky in
equatorial coordinates. The solid circles represent the E and S0 galaxies, while
triangles---the spiral galaxies.}
\end{figure}

2000.0; (3) morphological type of the galaxy in the de Vaucouleurs scale; (4)
radial velocities in km/s with respect to the centroid of the Local Group (LG);
(5) radial velocity measurement errors in km/s; (6) apparent B-magnitudes, taken
from the NED or estimated by eye at the lack of data in the NED; (7) projected
separation of the companion in kpc, determined from the radial velocity $V_{LG}$
in the standard model with the Hubble parameter $H_0$ = 73 km/s/Mpc; (8) absolute
magnitude of the 2MIG galaxy, corrected for the absorption in the Galaxy from
Schlegel et al. [21], (9) the logarithm of orbital mass of the ``2MIG galaxy---
companion'' pair in solar masses, calculated according to [22] as
$M_{orb}=(16/\pi G)\Delta V^2R_p,$ , where $G$ is the gravitational constant;
(10,11) the logarithm of the orbital mass-to-luminosity ratio in solar units,
in $B$ and $K_s$ -bands, respectively.

The distribution of 125 2MIG-galaxies
with faint companions is presented in equatorial coordinates in Fig. 1. Elliptical
and lenticular 2MIG galaxies ($T\leq0$) are marked by solid circles, while the
spiral galaxies are marked by triangles. Compared with the general distribution
of 3227 catalog galaxies, the 2MIG galaxies with companions reveal a more
heterogeneous distribution in the sky, which is to a large extent caused by
the presence of the Galactic absorption belt, as well as the geometric features
of the SDSS and other spectral surveys. This map reveals some evidence of
association of 2MIG galaxies with companions into multiple systems.

It should
be noted that the relative number of 2MIG galaxies with faint companions is
less than
4\%. If we assume that the main contribution to this number is given by the
SDSS survey, then the total relative number of such galaxies in the 2MIG catalog
may make up about 15\%, indicating a fairly high efficiency of the applied isolation
criterion.

\bigskip
{\bf \large 4. SOME PROPERTIES OF ISOLATED GALAXY---COMPANION PAIRS}
\bigskip

 The distribution of 214 2MIG galaxy---companion pairs by the radial velocity difference
and projected linear separation is presented in Fig. 2. Spiral 2MIG galaxies are marked
with triangles, and E and S0 galaxies -- with circles. Vertical bars indicate the mean
square error of the velocity difference. Radial velocities of the companions relative to
the main galaxy are distributed quite symmetrically with the mean of
$\langle V_{sat}-V_{2MIG}\rangle =+11\pm13$ km/s, and a standard deviation of 187 km/s.

The velocity dispersion of companions to elliptical galaxies is somewhat greater than
that of the spiral galaxies. The medians of the difference modulus are 205 and
130 km/s, respectively. A small contribution in this discrepancy is made by higher values
of the velocity measurement errors in the E and S0 galaxies. The nature of companion
distribution by the projected separations $R_p$ looks almost uniform in the range of
0--500 kpc, varying distinctly from the dependence $N(R_p)\propto R_p^{-1.7}$, obtained
in [7]. This may indicate the presence of a significant fraction of fictitious
companions at the separations of $R_p > 300$ kpc, which

\begin{figure}
\includegraphics{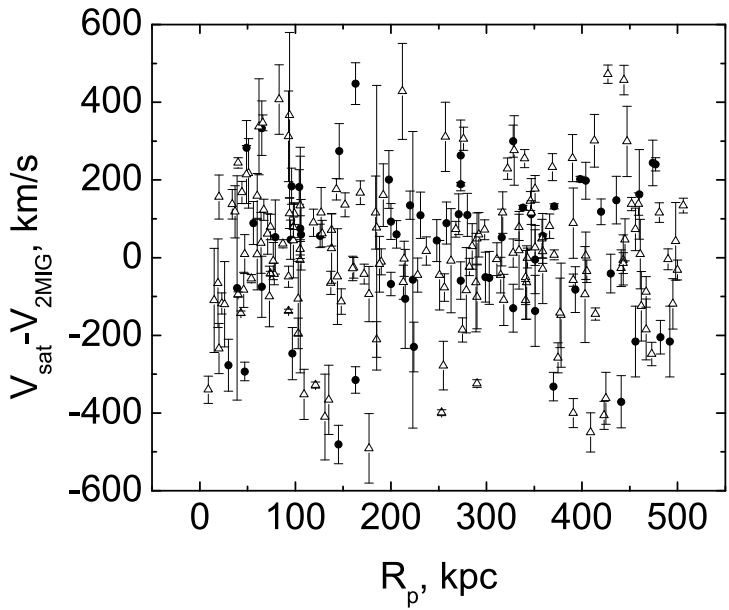}
\caption{ The distribution of 2MIG galaxies with companions by the difference of
radial velocities and projected linear separations. 2MIG galaxies of early types
are marked with circles, spiral galaxies -- with triangles.}
\end{figure}

were formed by the members of diffuse filaments and other elements of the large-scale
structure.

Figure 3 shows the relation between the apparent B-magnitude difference of
the given companion and the main galaxy, and the projected separation between them.
As expected, there is a noticeable trend of the decreasing difference of values with
increasing projected separation of the companion due to the 2MIG galaxy isolation
selection conditions. On the average, the companions are approximately ~3.5$^m$,
or 25 times fainter than the 2MIG galaxies, and this difference is somewhat greater
for the E and S0 galaxies (circles) than for the spiral ones (triangles).

If the 2MIG
galaxy companions are subject to Keplerian motions, their relative radial velocities
should correlate with the luminosity of the main component of the pair. Figure 4 presents
the distribution of 2MIG galaxy--companion pairs by the radial velocity difference
modulus and the absolute B-magnitude of the main galaxy, separately for the companions
around the elliptical (E, S0) and spiral 2MIG galaxies. The slope of linear regressions
in the figure shows the presence of a weak expected trend for the E and S0 galaxies,
and a lack of a significant correlation for spiral galaxies. Note that the overwhelming
majority of 2MIG galaxies with companions are the objects of high luminosity with the
median absolute magnitudes of --20.3 (S) and --20.4 (E, S0).

An important
dynamic characteristic of 2MIG galaxies with companions is an estimate of
their orbital mass

$\langle M_{orb}\mid e\rangle=(32/3\pi G)(1-2e^2/3)^{-1}\Delta V^2R_p$,
where $e$ is the eccentricity of companion's orbit [23]. Following [22],
for an ensemble of pairs we have adopted as an average the value of
$\langle e^2\rangle=1/2$. This estimate will be statistically unbiased
only at zero measurement errors $\sigma_V$ of galaxy radial velocities.
As we can see from column (5) of the Table, the  errors are quite
significant for many companions.

Therefore, in addition to $M_{orb}$ ,
we calculated for each pair an unbiased estimate of orbital mass
$M_{cor}=M_{orb} (\Delta V^2-\sigma^2_{V1}-\sigma^2_{V2})/\Delta V^2$,
making the quadratic subtraction of measurement errors from the
\begin{figure}
\includegraphics{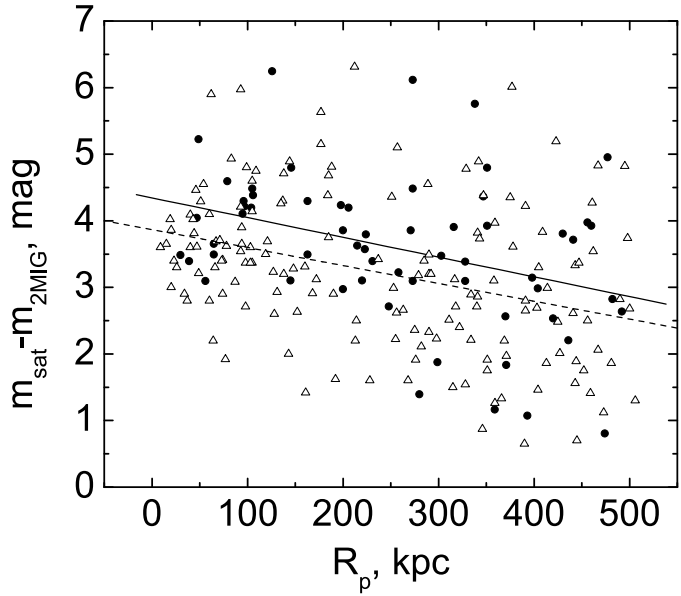}

\caption{ The distribution of 2MIG galaxies with companions by the
difference in apparent B-magnitudes and projected separations.}
\end{figure}

difference of radial velocities $\Delta V$. The distribution of 2MIG
galaxy--companion pairs by the $M_{orb}$ and $M_{cor}$ values is
demonstrated in Fig. 5 in logarithmic scale. Approximately 30\% of
all pairs have negative values of $M_{cor}$ due to large errors of
$\sigma_V$. It is clear that the role of  $\sigma_V$ errors is
significant in the region of $\log(M/M_{\odot})<12$. In general,
if we account for the velocity measurement errors for an ensemble of
2MIG companions with the mean square velocity difference of 187 km/s
and the mean square difference error of 62 km/s, it will reduce the mean
estimate of the orbital mass by 16\%.

The distribution of orbital mass
estimates, normalized for the blue ($B$-band) and infrared ($K_s$ -band)
luminosity is illustrated in Fig. 6. The subsample of E and S0 galaxies
in it is shaded. We have neglected the luminosity of 2MIG companion
galaxies. It follows from these data that the median ratio of the
orbital mass to the $B$-luminosity in the E and S0 galaxies is twice
higher than that in the spirals. In the $K_s$ -band this difference
increases up to 3.7. The reason for this is the abundance among the
2MIG galaxies of spiral
galaxies, seen nearly edge-on. Internal absorption of light in them,
which we have not considered, reaches $\Delta m_B\simeq1^m$. Therefore,
we believe that the 3--4-fold difference between the median values of
$M_{orb}/L_K$ for elliptical and spiral galaxies is more realistic.
However, this effect can also be caused by the photometric feature of
the 2MASS survey, which underestimates the luminosity of the peripheral
regions of galactic disks.

\bigskip
{\bf \large 5. DISCUSSION}
\bigskip

 The orbital mass-to-luminosity ratio for 2MIG galaxies with companions
(Fig. 6) is distributed over a fairly wide range, covering approximately
5 orders of magnitude. The shape of the left wing of the distribution is
affected by the projection factors, as well as by the anticipated nature
of the companions motion (the function of their orbit eccentricities). At
the right wing of the distribution, in the range of
$M_{orb}/L_K>100M_{\odot}/L_{\odot}$ values, an admixture of fictitious
2MIG galaxy--companion pairs is potentially noticeable. Recall that in the
standard $\Lambda$CDM model,
\begin{figure}
\includegraphics[scale=0.8]{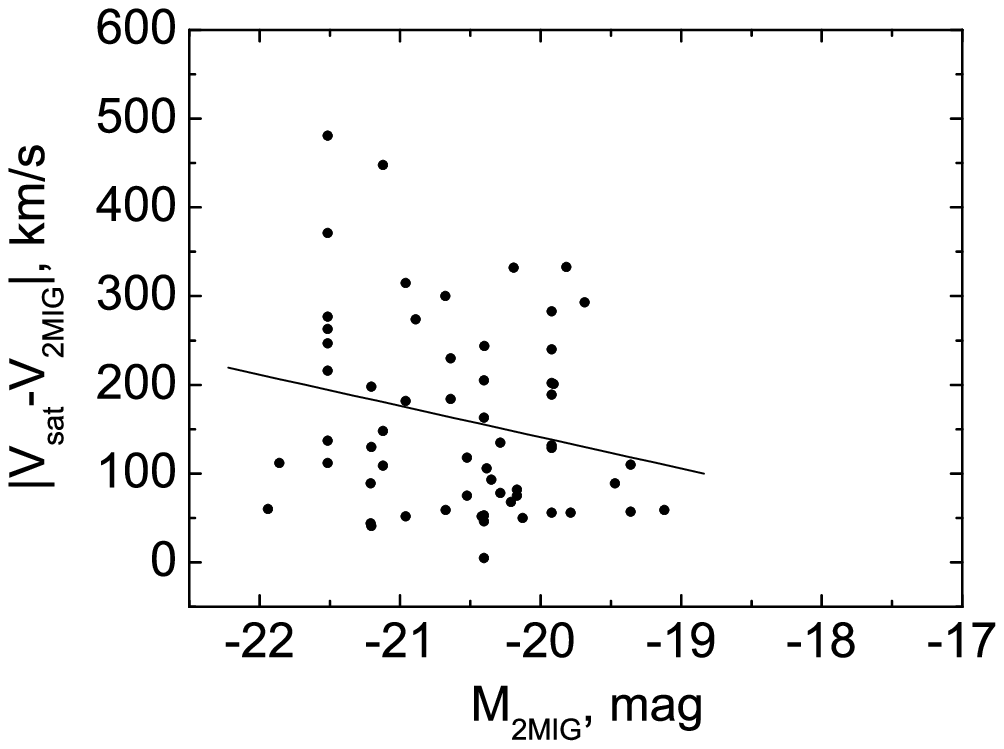}
\includegraphics[scale=0.8]{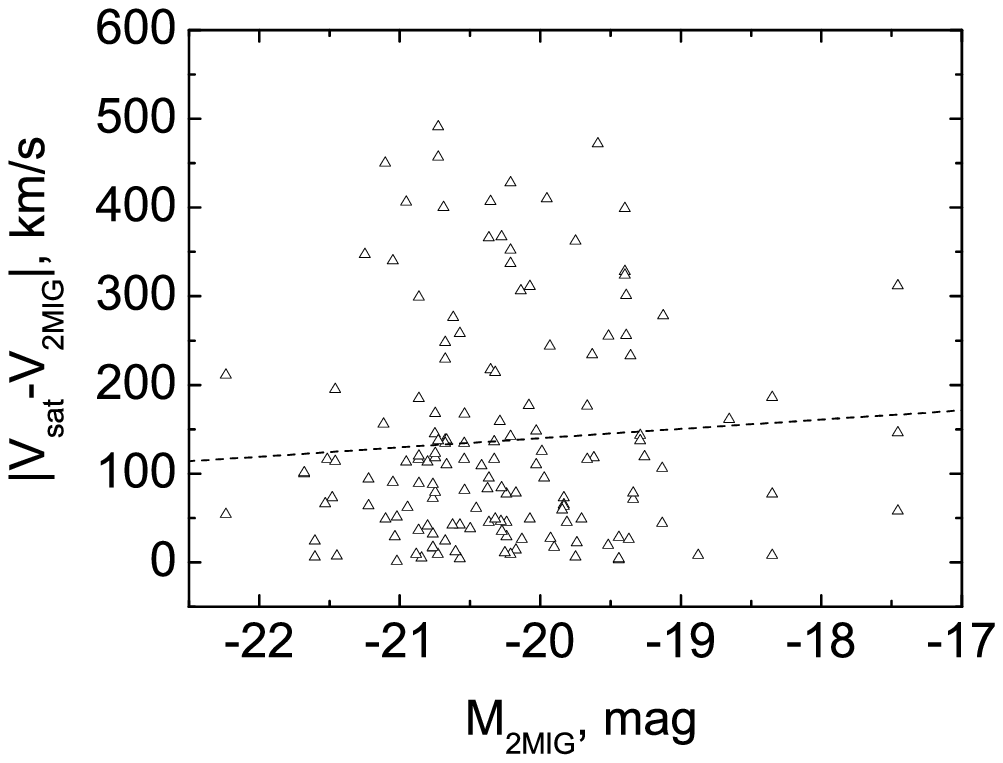}

\caption{ The module of radial velocity difference of a given 2MIG galaxy
and its companion, depending on the absolute magnitude of the 2MIG-galaxy.
The E and S0 galaxies are located in the left plot, and spiral galaxies ---
to the right.}
\end{figure}

\topmargin=-1cm
\begin{figure}
\includegraphics{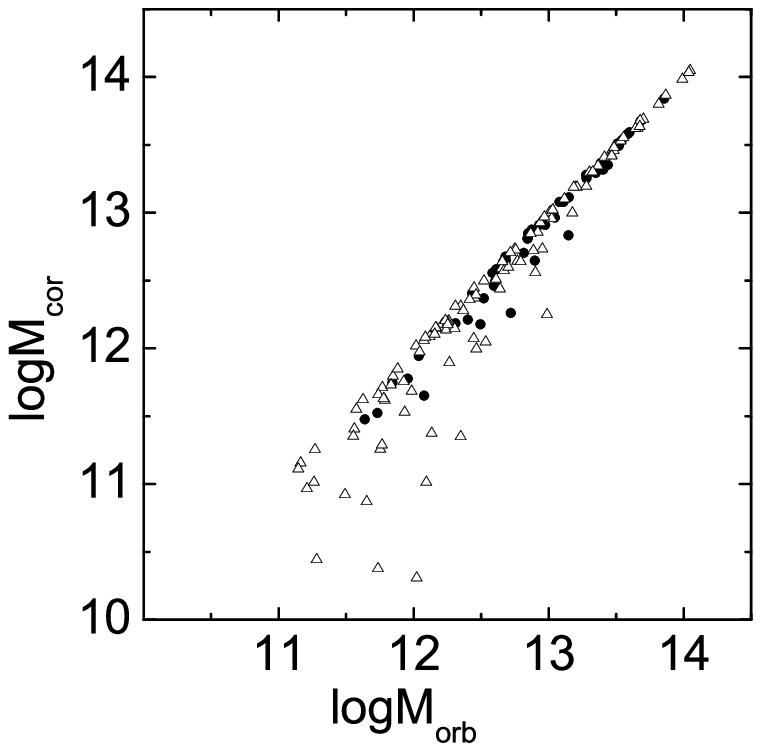}
\includegraphics{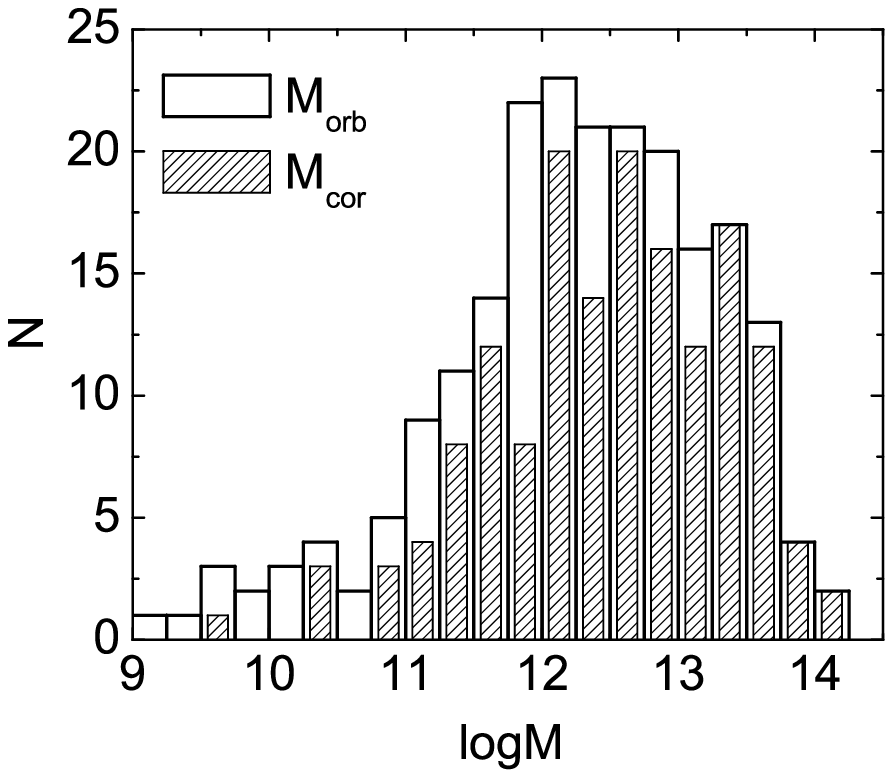}
\caption{ Left plot: the relation between the biased and unbiased estimates
of the orbital mass of 2MIG galaxy--companion pairs. Right plot: the
distribution of the number of 2MIG galaxy--companion pairs by the biased
and unbiased (hatched)
 orbital mass estimates.} 
\end{figure}

\topmargin=-1cm
\begin{figure}
\includegraphics[scale=0.9]{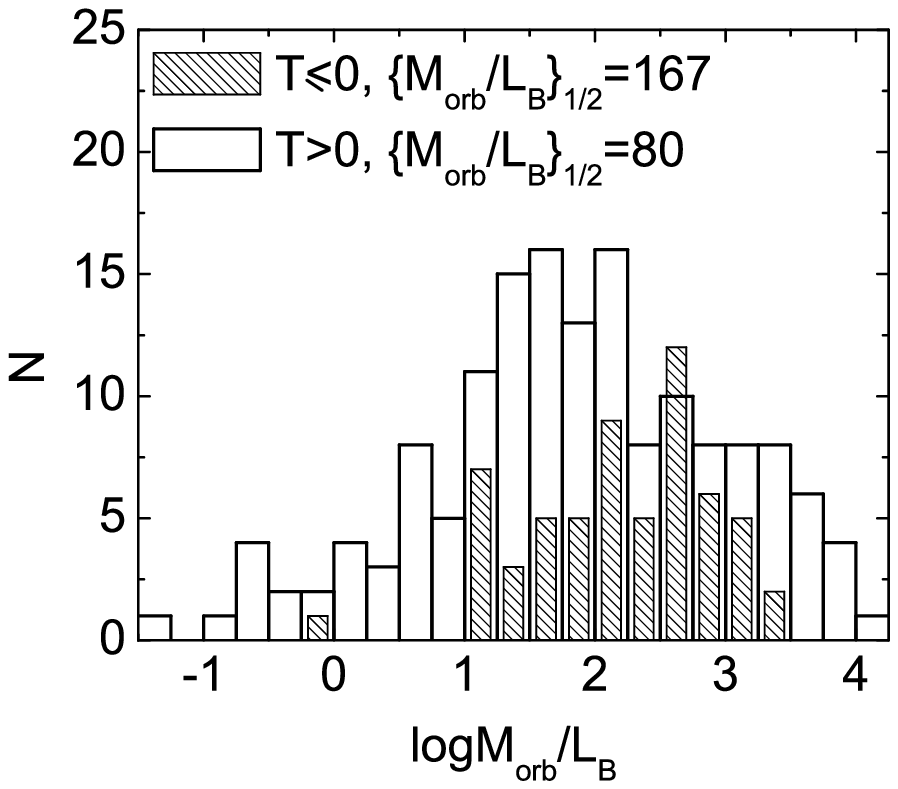}
\includegraphics[scale=0.9]{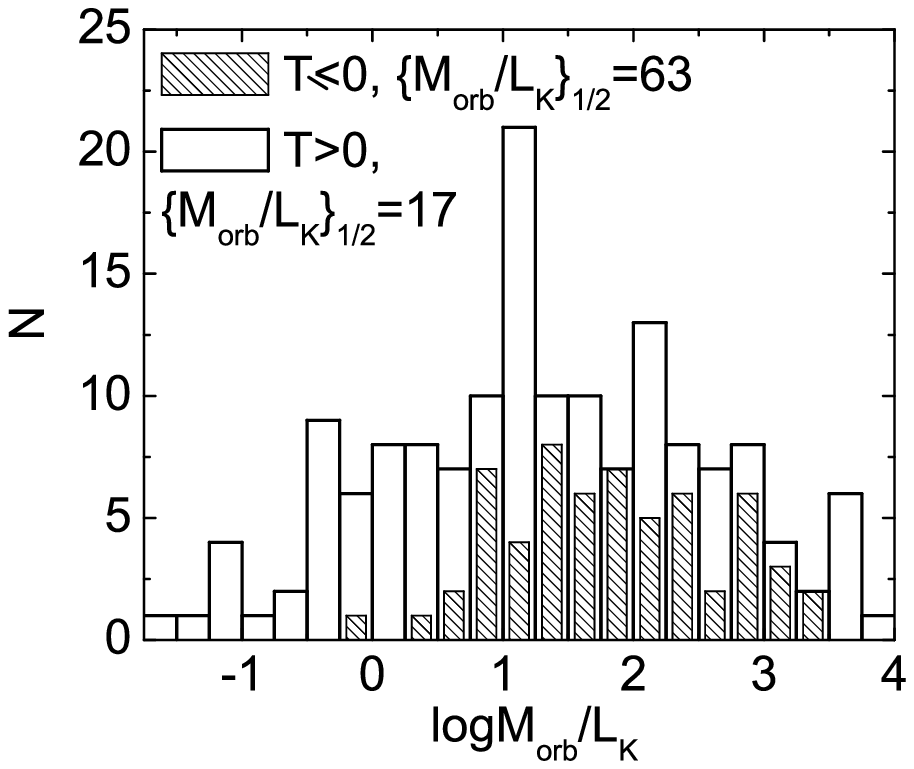}
\caption{ The distribution of 2MIG galaxies with companions by logarithm
of the orbital mass-to-luminosity ratio in the B (left) and K (right)
bands. 2MIG galaxies of early types are shaded.}
\end{figure}

the value of the mean cosmic density $\Omega_m=0.27$ corresponds to
$M/L_K\simeq100 M_{\odot}/L_{\odot}$. The statistics of virial masses
in the systems of galaxies of various scales in the near
($V_{LG} < 3500$ km/s) Universe reveals [24--26] that the median value of
$M_{vir}/L_K$ is  $11M_{\odot}/L_{\odot}$, $15M_{\odot}/L_{\odot}$ and
$31M_{\odot}/L_{\odot}$ for pairs, triplets, and groups of galaxies,
respectively, which is significantly below the mean global ratio of about
$\sim100M_{\odot}/L_{\odot}$. The median
$M_{orb}/L_K=28M_{\odot}/L_{\odot}$ in our sample is almost identical with
the median for groups of galaxies.

As we have already noted, the decrease
in the number of 2MIG galaxy--companion pairs with the projected separation
$R_p$ happens much slower than  should by law, about $R_p^{-1.7}$ . The
analysis of data from the Table and Fig. 1 reveals the presence of
associations of 2MIG galaxies with companions, having similar positions
in the sky and close radial velocities. The examples of such associations
may be 2MIG 2 and 2MIG 13, 2MIG 239 and 243, 2MIG 1871 and 1873, 2MIG 1987
and 1997. Moreover, in the case of a system of companions around 2MIG 274
and 282, there exists an object LCRS B021927.3--414244, which falls in the
interval of $\Delta V<500$ km/s and $R_p < 500$ kpc with respect to both
2MIG galaxies. The statistics of 2MIG galaxies with $k\geq 5$ number of
companions can serve as indirect evidence of the presence of scattered
associations. The Table has four such cases, whereas their expected number
at the Poisson distribution should be several orders of magnitude smaller.

The presence of diffuse components of the large-scale structure of the
universe (filaments, walls), even in the regions of very low matter density,
does apparently make the problem of separation of physical and spurious
companions of isolated galaxies very difficult. On the other hand, an
example of a nearby isolated galaxy NGC 1156 and its dwarf companion AGES
J030039+254656 indicates that the estimates of the orbital
mass-to-K-luminosity ratio in these indisputably isolated pairs can reach
values of around $\sim200M_{\odot}/L_{\odot}$.

\bigskip
{\bf \large 6. CONCLUDING REMARKS}
\bigskip

 Analyzing the data on radial velocities and projected separations of
galaxies in the vicinity of 2MIG catalog objects, we conclude on a good
dynamic isolation of the majority of catalog galaxies. According to our
estimates, not more than 15\% of 2MIG galaxies have small companions in
their vicinities with relative velocities of  $\Delta V<500$ km/s and
projected separations of $R_p < 500$ kpc. At the characteristic difference
of apparent magnitudes of about $3.5^m$, these companions have little
influence on the evolution of 2MIG galaxies.

The median ratio of the
orbital mass to the K-luminosity in isolated galaxies is
$17M_{\odot}/L_{\odot}$ for
spirals, and $63M_{\odot}/L_{\odot}$ for the E and S0 galaxies. These
values can be somewhat inflated due to the presence of false companions
from diffuse associations and filaments located in the regions of low
galaxy number density in the 2MIG sample. The advent of new data from
the surveys of galaxy radial velocities, carried out in the optical (SDSS)
and radio (ALFALFA) ranges will certainly allow to explore the features of
2MIG galaxies in more detail.

{\bf ACKNOWLEDGMENTS}

 This study was made owing to the support of the following grants: the grant
of the Russian Foundation for Basic Research (RFBR) (project no.
11--02--90449--Ukr--f--a), the grant of the State Foundation for Basic
Research of the Ukraine no. F40.2/49, as well as the Cosmomicrophysics
program of the National Academy of Sciences of Ukraine. We made use of
the NED database (http://nedwww. ipac.caltech.edu), as well as the digital
sky surveys DSS (http://archive.eso.org/dss/dss), and SDSS
(http://www.sdss.org).

{\bf  REFERENCES}

1. D. Zaritsky, R. Smith, C. Frenk, and S. D. M. White, Astrophys. J. 405, 464 (1993).

2. S. Herbert-Fort, D. Zaritsky, Y. J. Kim, et al., Monthly Notices Roy. Astronom. Soc. 384, 803 (2008). 

3. K. N. Abazajian, J. K. Adelman-McCarthy, M. A. Agueros, et al., Astrophys. J. Suppl. 182, 543 (2009). 

4. M. Colless et al., Monthly Notices Roy. Astronom. Soc. 328, 1039 (2001). 

5. J. P. Huchra, L. M. Macri, K. L. Masters, et al., arXiv: 1108.0669. 

6. F. Prada, M. Vitvitska, A. Klypin, et al., Astrophys. J. 598, 260 (2003). 

7. J. Chen, A. V. Kravtsov, F. Prada, et al., Astrophys. J. 647, 86 (2006). 

8. C. Conroy, F. Prada, J. A. Newman, et al., Astrophys. J. 654, 153 (2007).

9. P. Norberg, C. S. Frenk, and S. Cole, Monthly Notices Roy. Astronom. Soc. 383, 646 (2008). 

10. H. S. Hwang and C. Park, Astrophys. J. 720, 522 (2010). 

11. L. R. Jones, T. J. Ponman, A. Horton, et al., Monthly Notices Roy. Astronom. Soc. 343, 627 (2003). 

12. A. M. von Benda-Beckmann, E. D'Onghia, S. Gottloeber, et al., Monthly Notices Roy. Astronom. Soc. 386, 2345 (2008). 

13. R. N. Proctor, C. Mendes de Oliveira, R. Dupke, et al., arXiv: 1108.1349. 

14. V. E. Karachentseva, S. N. Mitronova, O. V. Melnyk, and I. D. Karachentsev, Astrophysical Bulletin 65, 1 (2010).

 15. V. E. Karachentseva, Soobscheniya SAO 8, 3 (1973).

16. S. Verley, S. Leon, L. Verdes-Montenegro, et al., Astronom. and Astrophys. 472, 121 (2007). 

17. M. F. Skrutskie, S. E. Schneider, R. Steining, et al., ASSL 210, 25 (1997).

 18. T. N. Jarrett, T. Chester, R. M. Cutri, et al., Astronom. J. 119, 2498 (2000).

 19. Yu. N. Kudrya, V. E. Karachentseva, and I. D. Karachentsev, Astrofizika 54, 45 (2011) .

20. R. F. Minchin, E. Momjian, R. Auld, et al., Astronom. J. 140, 1093 (2010). 

21. D. J. Schlegel, D. P. Finkbeier, and M. Davis, Astrophys. J. 500, 525 (1998). 

22. J. Heisler, S. Tremaine, and J. N. Bahcall, Astrophys. J. 298, 8 (1985). 

23. I. D. Karachentsev, Binary Galaxies (Moscow, Nauka, 1987) [in Russian]. 

24. I. D. Karachentsev and D. I. Makarov, Astrophysical Bulletin 63, 299 (2008).

 25. D. I. Makarov and I. D. Karachentsev, Astrophysical Bulletin 64, 24 (2009).

 26. D. I. Makarov and I. D. Karachentsev, Monthly Notices Roy. Astronom. Soc. 412, 2498 (2011).

\clearpage
\topmargin=-3cm
{\small

\renewcommand{\baselinestretch}{0.8}
\begin{table}
\caption{A list of 2MIG galaxies and their companions with $\Delta V< 500$
km/s and $R_p < 500$ kpc}
\begin{tabular}{rccrrrrrrrr} \hline \\
2MIG       &       RA  Dec.   &       T       &       $V_{LG}$        &       $\sigma_{V}$    &       $m_{B}$ &       $R_{p}$ &       $M_{B}^{c}$     &       log$M_{orb}$    &       log$(M_{orb}/L_B)$       &       log$(M_{orb}/L_K)$       \\
\hline
1       &       2       &       3       &       4       &       5       &       6       &       7       &       8       &       9       &       10      &       11      \\
\hline
2	&	000058.3-333639	&	5	&	6942	&	10	&	14.72	&		&	-20.24	&		&		&		\\
sat	&	000136.5-334125	&		&	6865	&	34	&	17.33	&	256	&	-17.60	&	12.24	&	1.99	&	1.59	\\
sat	&	000132.8-334532	&		&	6897	&	45	&	16.21	&	315	&	-18.73	&	11.86	&	1.60	&	1.21	\\
sat	&	000003.3-333035	&		&	6913	&	89	&	18.68	&	359	&	-16.26	&	11.58	&	1.32	&	0.93	\\
13	&	000834.5-335130	&	3	&	6823	&	10	&	13.3	&		&	-21.60	&		&		&		\\
sat	&	000829.7-335628 	&		&	6847	&	89	&	18.02	&	138	&	-16.90	&	10.97	&	0.17	&	-0.49	\\
sat	&	000945.6-334438	&		&	6817	&	64	&	16.64	&	443	&	-18.27	&	10.27	&	-0.53	&	-1.19	\\
48	&	002529.9+455518	&	3	&	5353	&	28	&	14.68	&		&	-19.93	&		&		&		\\
sat	&	002701.9+454214	&		&	5326	&	38	&	17.3	&	441	&	-17.30	&	10.97	&	0.84	&	-0.23	\\
62	&	003040.4-284245	&	1	&	7331	&	100	&	15.06	&		&	-20.03	&		&		&		\\
sat	&	003038.4-284301	&		&	7221	&	89	&	18.71	&	15	&	-16.35	&	11.40	&	1.23	&	0.48	\\
sat	&	003017.0-283052	&		&	7183	&	89	&	18.67	&	378	&	-16.37	&	12.99	&	2.82	&	2.07	\\
64	&	003135.8+143644	&	6	&	11641	&	16	&	15.53	&		&	-20.76	&		&		&		\\
sat	&	003123.9+143059	&		&	11713	&	20	&	17.8	&	298	&	-18.55	&	12.26	&	1.80	&	1.15	\\
sat	&	003143.0+142911	&		&	11657	&	26	&	16.8	&	359	&	-19.51	&	11.04	&	0.57	&	-0.08	\\
sat	&	003157.9+142722	&		&	11609	&	20	&	18.2	&	500	&	-18.08	&	11.75	&	1.29	&	0.64	\\
75	&	003755.6+045441	&	5	&	8659	&	13	&	15.42	&		&	-20.08	&		&		&		\\
sat	&	003836.3+045348	&		&	8836	&	32	&	17.18	&	351	&	-18.37	&	13.12	&	2.93	&	2.16	\\
77	&	003823.7+150222	&	1	&	5592	&	27	&	14.9	&		&	-19.90	&		&		&		\\
sat	&	003748.5+145558	&		&	5609	&	24	&	18.3	&	237	&	-16.49	&	10.85	&	0.73	&	0.25	\\
78	&	003839.9+172410	&	3	&	5643	&	15	&	14.8	&		&	-19.83	&		&		&		\\
sat	&	003750.3+172259	&		&	5716	&	14	&	16.4	&	268	&	-18.26	&	12.21	&	2.12	&	1.68	\\
85	&	004325.7-501058	&	3	&	8523	&	9	&	13.14	&		&	-22.24	&		&		&		\\
sat	&	004327.3-500924	&		&	8469	&	5	&	17.69	&	54	&	-17.67	&	11.27	&	0.21	&	-0.15	\\
sat	&	004300.1-501432	&		&	8312	&	45	&	16.89	&	185	&	-18.43	&	12.99	&	1.93	&	1.57	\\
90	&	004712.8+290811	&	2	&	5948	&	33	&	15.95	&		&	-18.88	&		&		&		\\
sat	&	004632.9+290118	&		&	5940	&	130	&	18.6	&	263	&	-16.21	&	10.30	&	0.59	&	-0.30	\\
108	&	005502.4-134041	&	-2	&	6375	&	36	&	15.03	&		&	-19.79	&		&		&		\\
sat	&	005510.1-133031	&		&	6431	&	45	&	16.2	&	359	&	-18.63	&	12.12	&	2.05	&	1.37	\\
110	&	005652.4-315747	&	0	&	5611	&	38	&	14.59	&		&	-19.91	&		&		&		\\
sat	&	005720.4-315113	&		&	5812	&	64	&	18.83	&	198	&	-15.75	&	12.97	&	2.85	&	2.21	\\
158	&	012853.2+134738	&	5	&	6547	&	17	&	14.8	&		&	-20.17	&		&		&		\\
sat	&	012906.4+134539	&		&	6625	&	20	&	17.5	&	98	&	-17.49	&	11.85	&	1.62	&	1.18	\\
sat	&	012855.8+140436	&		&	6533	&	21	&	16.3	&	443	&	-18.61	&	11.07	&	0.84	&	0.40	\\
159	&	012902.9+321956	&	0	&	5622	&	34	&	14.49	&		&	-20.19	&		&		&		\\
sat	&	012920.8+320352	&		&	5290	&	14	&	17	&	370	&	-17.49	&	13.68	&	3.44	&	2.89	\\
160	&	013029.1-224003	&	5	&	1645	&	3	&	11.44	&		&	-20.37	&		&		&		\\
sat	&	013011.9-224545	&		&	1562	&	45	&	15.91	&	46	&	-15.80	&	11.56	&	1.25	&	1.14	\\
165	&	013142.1-005600	&	2	&	5542	&	20	&	14	&		&	-20.57	&		&		&		\\
sat	&	013128.0-005603	&		&	5500	&	19	&	17.6	&	78	&	-16.94	&	11.21	&	0.82	&	-0.08	\\
sat	&	013154.6-004853	&		&	5500	&	16	&	17.1	&	172	&	-17.44	&	11.55	&	1.17	&	0.27	\\
sat	&	013050.5-004459	&		&	5284	&	33	&	18.3	&	375	&	-16.12	&	13.47	&	3.08	&	2.18	\\
sat	&	013028.1-004348	&		&	5538	&	18	&	16.77	&	490	&	-17.75	&	10.16	&	-0.23	&	-1.13	\\
186	&	014035.7-333717	&	0	&	8791	&	34	&	15.1	&		&	-20.38	&		&		&		\\
sat	&	014044.0-333123	&		&	8685	&	123	&	18.73	&	215	&	-16.73	&	12.46	&	2.15	&	1.32	\\
194	&	014314.2+085322	&	3	&	5677	&	11	&	13.76	&		&	-20.95	&		&		&		\\
sat	&	014253.4+084920	&		&	5564	&	31	&	17	&	148	&	-17.63	&	12.35	&	1.81	&	1.21	\\
sat	&	014354.3+083729	&		&	5271	&	34	&	18.9	&	423	&	-15.60	&	12.65	&	2.11	&	1.52	\\
212	&	015541.1-295520	&	0	&	4327	&	14	&	13.41	&		&	-20.52	&		&		&		\\
sat	&	015557.7-300014	&		&	4402	&	30	&	17.91	&	105	&	-16.07	&	11.84	&	1.47	&	0.73	\\
sat	&	015524.3-301926	&		&	4445	&	30	&	15.96	&	420	&	-18.04	&	12.85	&	2.48	&	1.73	\\
222	&	015808.5+020352	&	6	&	6393	&	12	&	14.91	&		&	-19.93	&		&		&		\\
sat	&	015804.7+020237	&		&	6637	&	6	&	19	&	40	&	-15.92	&	12.45	&	2.31	&	1.76	\\
226	&	020008.9+123922	&	2	&	3623	&	22	&	14.03	&		&	-19.74	&		&		&		\\
sat	&	020001.9+123218	&		&	3645	&	10	&	18.2	&	105	&	-15.61	&	10.78	&	0.72	&	0.09	\\
\end{tabular}
\end{table}

\begin{table}
\begin{tabular}{lccrrrrrrrr} \hline \\
\hline
1       &       2       &       3       &       4       &       5       &       6       &       7       &       8       &       9       &       10      &       11      \\
\hline
239	&	020540.3-004141	&	5	&	12904	&	13	&	15.7	&		&	-20.68	&		&		&		\\
sat	&	020519.5-004325	&		&	12880	&	18	&	17.8	&	282	&	-18.56	&	11.28	&	0.85	&	0.10	\\
sat	&	020600.1-004531	&		&	13133	&	24	&	18.1	&	322	&	-18.32	&	13.30	&	2.87	&	2.12	\\
sat	&	020504.8-004242	&		&	13042	&	26	&	17.1	&	459	&	-19.29	&	13.01	&	2.58	&	1.83	\\
sat	&	020525.0-003320	&		&	12656	&	27	&	16.8	&	473	&	-19.51	&	13.54	&	3.10	&	2.36	\\
243	&	020616.0-001729	&	0	&	12807	&	30	&	14.4	&		&	-21.94	&		&		&		\\
sat	&	020619.1-002126	&		&	12867	&	19	&	18.6	&	206	&	-17.75	&	11.96	&	1.02	&	0.42	\\
253	&	021006.1-325623	&	1	&	3293	&	10	&	13.12	&		&	-20.21	&		&		&		\\
sat	&	021028.0-325514	&		&	3630	&	123	&	19.02	&	62	&	-14.52	&	12.92	&	2.68	&	2.13	\\
sat	&	021007.5-324806	&		&	2941	&	64	&	17.87	&	109	&	-15.22	&	13.20	&	2.96	&	2.41	\\
sat	&	021044.7-324222	&		&	3721	&	123	&	19.43	&	212	&	-14.17	&	13.66	&	3.42	&	2.87	\\
sat	&	020807.1-331043	&		&	3151	&	89	&	19.13	&	377	&	-14.11	&	12.95	&	2.71	&	2.16	\\
sat	&	020720.5-330157	&		&	3302	&	89	&	17.38	&	461	&	-15.95	&	10.54	&	0.30	&	-0.25	\\
274	&	021958.6-412411	&	3	&	4952	&	13	&	13.55	&		&	-20.67	&		&		&		\\
sat	&	022126.9-412905	&		&	4842	&	47	&	16.4	&	341	&	-17.76	&	12.67	&	2.24	&	1.73	\\
sat	&	021906.9-414458	&		&	5090	&	11	&	15.3	&	452	&	-18.98	&	13.01	&	2.59	&	2.07	\\
282	&	022247.3-412216	&	-2	&	4972	&	40	&	13.02	&		&	-21.21	&		&		&		\\
sat	&	022126.9-412905	&		&	4842	&	47	&	16.4	&	328	&	-17.76	&	12.82	&	2.17	&	1.82	\\
sat	&	022435.9-412306	&		&	5170	&	26	&	16.01	&	404	&	-18.30	&	13.28	&	2.64	&	2.29	\\
sat	&	022438.1-412832	&		&	4931	&	30	&	16.83	&	430	&	-17.38	&	11.89	&	1.24	&	0.89	\\
285	&	022528.3-253817	&	4	&	4808	&	4	&	14.09	&		&	-20.07	&		&		&		\\
sat	&	022505.8-253246	&		&	4759	&	123	&	18.97	&	144	&	-15.16	&	11.61	&	1.42	&	1.34	\\
sat	&	022554.4-252613	&		&	5119	&	89	&	19.17	&	257	&	-15.11	&	13.46	&	3.28	&	3.20	\\
295	&	023036.2-313546	&	6	&	4566	&	37	&	14.93	&		&	-19.13	&		&		&		\\
sat	&	023011.0-313628	&		&	4610	&	89	&	18.3	&	98	&	-15.78	&	11.35	&	1.54	&	1.20	\\
sat	&	023009.7-313619	&		&	4460	&	89	&	18.53	&	103	&	-15.48	&	12.14	&	2.32	&	1.98	\\
319	&	024222.6-301920	&	-2	&	6435	&	19	&	14.16	&		&	-20.68	&		&		&		\\
sat	&	024240.0-302103	&		&	6494	&	89	&	18.54	&	106	&	-16.31	&	11.64	&	1.21	&	0.75	\\
331	&	024816.1+342511	&	2	&	5462	&	12	&	13.99	&		&	-20.75	&		&		&		\\
sat	&	024822.4+340612	&		&	5317	&	10	&	17	&	414	&	-17.70	&	13.01	&	2.55	&	2.14	\\
340	&	025158.8-332025	&	3	&	6310	&	10	&	14.45	&		&	-20.35	&		&		&		\\
sat	&	025202.2-332219	&		&	6527	&	89	&	18.74	&	51	&	-16.14	&	12.45	&	2.15	&	1.32	\\
sat	&	025155.6-331710	&		&	6717	&	89	&	19.38	&	83	&	-15.56	&	13.21	&	2.91	&	2.07	\\
394	&	031500.8-304229	&	3	&	4541	&	8	&	14.4	&		&	-19.63	&		&		&		\\
sat	&	031505.8-304247	&		&	4307	&	64	&	18.27	&	20	&	-15.65	&	12.12	&	2.11	&	1.38	\\
400	&	031713.2-323433	&	5	&	4431	&	7	&	13.26	&		&	-20.73	&		&		&		\\
sat	&	031653.7-324343	&		&	3940	&	89	&	18.88	&	177	&	-14.84	&	13.70	&	3.25	&	2.62	\\
407	&	032040.8-072340	&	-2	&	5462	&	24	&	14.4	&		&	-20.21	&		&		&		\\
sat	&	032106.1-071656	&		&	5394	&	18	&	17.4	&	200	&	-17.20	&	12.04	&	1.79	&	1.33	\\
516	&	041400.6+365052	&	3	&	6148	&	9	&	15.3	&		&	-21.45	&		&		&		\\
sat	&	041246.4+365350	&		&	6155	&	9	&	17	&	371	&	-19.48	&	10.20	&	-0.54	&	-0.98	\\
563	&	042849.1-445145	&	-2	&	4372	&	46	&	14.14	&		&	-19.82	&		&		&		\\
sat	&	042850.9-444802	&		&	4705	&	53	&	17.79	&	65	&	-16.32	&	12.93	&	2.84	&	2.26	\\
710	&	052644.5-191235	&	5	&	8108	&	13	&	14.76	&		&	-20.73	&		&		&		\\
sat	&	052613.0-192440	&		&	8099	&	46	&	16.7	&	77	&	-18.80	&	9.87	&	-0.58	&	-1.40	\\
777	&	055825.4+682740	&	3	&	4287	&	15	&	13.68	&		&	-20.60	&		&		&		\\
sat	&	055646.0+684436	&		&	4299	&	33	&	15.2	&	328	&	-19.07	&	10.67	&	0.27	&	-0.20	\\
1007	&	073737.1+415649	&	5	&	5917	&	6	&	14.8	&		&	-19.97	&		&		&		\\
sat	&	073729.8+415550	&		&	5822	&	3	&	18.4	&	40	&	-16.33	&	11.62	&	1.48	&	1.09	\\
1008	&	073836.5+373801	&	6	&	3892	&	7	&	14.5	&		&	-19.40	&		&		&		\\
sat	&	073913.7+374037	&		&	3564	&	3	&	18.2	&	121	&	-15.52	&	13.19	&	3.27	&	3.13	\\
sat	&	073951.9+374916	&		&	3568	&	7	&	18	&	290	&	-15.71	&	13.56	&	3.64	&	3.50	\\
sat	&	073934.5+380141	&		&	3493	&	3	&	17.5	&	253	&	-16.17	&	13.68	&	3.76	&	3.62	\\
1082	&	080448.1+204138	&	5	&	9260	&	9	&	14.2	&		&	-21.48	&		&		&		\\
sat	&	080420.8+205213	&		&	9333	&	54	&	16.7	&	456	&	-19.00	&	12.45	&	1.69	&	1.50	\\
1093	&	081015.2+335724	&	5	&	5219	&	4	&	13.7	&		&	-20.80	&		&		&		\\
sat	&	081025.2+340016	&		&	5178	&	4	&	16.6	&	74	&	-17.89	&	11.16	&	0.68	&	0.06	\\
sat	&	081021.1+340159	&		&	5332	&	27	&	18.5	&	99	&	-16.05	&	12.17	&	1.69	&	1.07	\\
\end{tabular}
\end{table}

\begin{table}
\begin{tabular}{lccrrrrrrrr} \hline \\
\hline
1       &       2       &       3       &       4       &       5       &       6       &       7       &       8       &       9       &       10      &       11      \\
\hline

1098	&	081406.8+235159	&	0	&	5961	&	20	&	14.5	&		&	-20.29	&		&		&		\\
sat	&	081403.8+235329	&		&	5883	&	288	&	17.9	&	39	&	-16.86	&	11.45	&	1.17	&	0.87	\\
sat	&	081407.3+234243	&		&	6096	&	30	&	17.6	&	220	&	-17.22	&	12.68	&	2.41	&	2.10	\\
1208	&	085832.8+281602	&	2	&	7927	&	7	&	14.2	&		&	-21.11	&		&		&		\\
sat	&	085831.1+281532	&		&	8083	&	56	&	17.2	&	20	&	-18.16	&	11.77	&	1.16	&	0.26	\\
1241	&	091458.3+512140	&	5	&	8298	&	22	&	14.6	&		&	-20.75	&		&		&		\\
sat	&	091448.1+512542	&		&	8180	&	16	&	18	&	23	&	-17.32	&	11.58	&	1.12	&	0.24	\\
sat	&	091409.0+512505	&		&	8466	&	18	&	18.4	&	44	&	-16.99	&	12.17	&	1.71	&	0.83	\\
sat	&	091618.3+512357	&		&	8421	&	16	&	18.3	&	67	&	-17.09	&	12.08	&	1.62	&	0.74	\\
sat	&	091329.8+511855	&		&	8377	&	24	&	18	&	74	&	-17.36	&	11.74	&	1.28	&	0.40	\\
1273	&	092908.2-023257	&	4	&	6745	&	39	&	14.14	&		&	-20.85	&		&		&		\\
sat	&	092945.3-022108	&		&	6750	&	47	&	15.6	&	404	&	-19.39	&	10.08	&	-0.42	&	-1.10	\\
1280	&	093011.7+555109	&	3	&	7614	&	12	&	14.7	&		&	-20.54	&		&		&		\\
sat	&	092948.1+554642	&		&	7781	&	28	&	17.6	&	168	&	-17.68	&	12.75	&	2.37	&	2.13	\\
sat	&	092956.4+553917	&		&	7695	&	29	&	16	&	366	&	-19.23	&	12.46	&	2.09	&	1.84	\\
1298	&	093652.5+374142	&	2	&	4343	&	10	&	14.6	&		&	-19.34	&		&		&		\\
sat	&	093617.0+373751	&		&	4414	&	40	&	17.8	&	138	&	-16.17	&	11.93	&	2.03	&	1.61	\\
sat	&	093801.6+375521	&		&	4421	&	32	&	17.5	&	334	&	-16.48	&	12.37	&	2.47	&	2.06	\\
1304	&	093823.4+433033	&	0	&	4435	&	48	&	14.5	&		&	-19.47	&		&		&		\\
sat	&	093837.9+432846	&		&	4524	&	28	&	17.6	&	56	&	-16.41	&	11.73	&	1.78	&	1.52	\\
1326	&	094541.5+045631	&	1	&	3550	&	15	&	13.85	&		&	-19.75	&		&		&		\\
sat	&	094555.9+050258	&		&	3544	&	290	&	17.2	&	105	&	-16.38	&	9.65	&	-0.41	&	-0.44	\\
sat	&	094738.6+044918	&		&	3188	&	65	&	16.3	&	425	&	-17.04	&	13.82	&	3.76	&	3.73	\\
1336	&	095056.2+621109	&	2	&	7476	&	23	&	14.9	&		&	-20.27	&		&		&		\\
sat	&	095242.1+621651	&		&	7441	&	17	&	18.2	&	405	&	-16.96	&	11.77	&	1.50	&	1.19	\\
1352	&	095449.5+091616	&	2	&	1283	&	11	&	13.04	&		&	-18.35	&		&		&		\\
sat	&	095407.3+092136	&		&	1291	&	90	&	17.1	&	60	&	-14.27	&	9.66	&	0.16	&	0.48	\\
sat	&	095430.5+095212	&		&	1360	&	366	&	17.7	&	185	&	-13.80	&	12.09	&	2.59	&	2.91	\\
sat	&	095529.7+082326	&		&	1097	&	29	&	15.4	&	275	&	-15.65	&	13.03	&	3.53	&	3.85	\\
1363	&	095929.5-224935	&	0	&	2105	&	17	&	12.85	&		&	-19.69	&		&		&		\\
sat	&	095919.6-224424	&		&	1812	&	17	&	16.9	&	47	&	-15.31	&	12.68	&	2.64	&	3.04	\\
1382	&	100453.9+050346	&	-2	&	3804	&	46	&	14.3	&		&	-19.36	&		&		&		\\
sat	&	100551.3+050022	&		&	3747	&	379	&	17.9	&	223	&	-15.75	&	11.93	&	2.03	&	1.70	\\
sat	&	100447.8+052210	&		&	3914	&	31	&	15.7	&	280	&	-18.02	&	12.59	&	2.69	&	2.36	\\
1394	&	100820.6+315146	&	3	&	5122	&	8	&	14	&		&	-20.33	&		&		&		\\
sat	&	100751.7+315247	&		&	5238	&	64	&	17.2	&	127	&	-17.15	&	12.31	&	2.01	&	1.21	\\
sat	&	100746.6+315338	&		&	5258	&	30	&	16.6	&	152	&	-17.75	&	12.52	&	2.23	&	1.42	\\
1439	&	102915.5+060741	&	6	&	3380	&	11	&	14.8	&		&	-18.66	&		&		&		\\
sat	&	102824.7+061419	&		&	3541	&	80	&	16.4	&	192	&	-17.14	&	12.77	&	3.15	&	2.54	\\
1465	&	104028.4+091057	&	0	&	5592	&	12	&	13.6	&		&	-20.96	&		&		&		\\
sat	&	104026.1+091537	&		&	5774	&	43	&	17.8	&	104	&	-16.83	&	12.61	&	2.07	&	2.21	\\
sat	&	104008.8+091629	&		&	5277	&	32	&	17.9	&	163	&	-16.53	&	13.28	&	2.74	&	2.88	\\
sat	&	104040.1+092449	&		&	5644	&	72	&	17.5	&	316	&	-17.07	&	12.00	&	1.46	&	1.60	\\
1488	&	104857.5-044538	&	5	&	7707	&	28	&	14.78	&		&	-20.54	&		&		&		\\
sat	&	104852.5-044849	&		&	7841	&	124	&	19.37	&	105	&	-15.98	&	12.35	&	1.97	&	1.41	\\
sat	&	104915.0-044132	&		&	7823	&	90	&	19.14	&	184	&	-16.19	&	12.46	&	2.09	&	1.53	\\
1495	&	105115.1+022616	&	1	&	15216	&	57	&	15.1	&		&	-21.68	&		&		&		\\
sat	&	105115.5+022728	&		&	15116	&	53	&	18.5	&	73	&	-18.26	&	11.93	&	1.10	&	0.56	\\
sat	&	105056.0+022600	&		&	15115	&	59	&	18.3	&	290	&	-18.47	&	12.53	&	1.70	&	1.16	\\
1502	&	105444.3-170231	&	5	&	3991	&	48	&	14.5	&		&	-19.39	&		&		&		\\
sat	&	105421.8-172626	&		&	4247	&	37	&	15.13	&	390	&	-18.87	&	13.48	&	3.57	&	2.92	\\
sat	&	105428.7-163649	&		&	4292	&	48	&	16.4	&	413	&	-17.68	&	13.64	&	3.73	&	3.08	\\
1507	&	105550.0+312332	&	3	&	10474	&	43	&	15.5	&		&	-20.42	&		&		&		\\
sat	&	105517.5+312639	&		&	10365	&	50	&	18.2	&	318	&	-17.68	&	12.64	&	2.31	&	0.97	\\
1512	&	105809.8-004629	&	5	&	6204	&	23	&	15.5	&		&	-19.36	&		&		&		\\
sat	&	105831.2-010025	&		&	6437	&	26	&	17.7	&	369	&	-17.24	&	11.56	&	1.66	&	0.74	\\
\end{tabular}
\end{table}

\begin{table}
\begin{tabular}{lccrrrrrrrr} \hline \\
\hline
1       &       2       &       3       &       4       &       5       &       6       &       7       &       8       &       9       &       10      &       11      \\
\hline
1524	&	110037.2+112455	&	2	&	8049	&	52	&	15	&		&	-20.29	&		&		&		\\
sat	&	110041.3+112320	&		&	8208	&	25	&	17.8	&	60	&	-17.53	&	12.25	&	1.98	&	1.00	\\
1538	&	110434.5+160342	&	0	&	6233	&	27	&	14.1	&		&	-20.64	&		&		&		\\
sat	&	110422.9+160624	&		&	6417	&	38	&	18.4	&	96	&	-16.40	&	12.58	&	2.17	&	2.07	\\
sat	&	110458.3+161040	&		&	6003	&	58	&	17.9	&	224	&	-16.76	&	13.15	&	2.73	&	2.64	\\
1539	&	110523.9-023142	&	1	&	8724	&	64	&	15.72	&		&	-19.95	&		&		&		\\
sat	&	110513.5-022851	&		&	8314	&	90	&	18.65	&	131	&	-16.92	&	12.79	&	2.65	&	2.06	\\
1543	&	110656.6+071026	&	6	&	1252	&	12	&	13.86	&		&	-17.45	&		&		&		\\
sat	&	110559.6+072225	&		&	1564	&	116	&	17.4	&	93	&	-14.40	&	13.02	&	3.88	&	3.64	\\
sat	&	111128.3+065427	&		&	1398	&	21	&	14.8	&	346	&	-16.82	&	12.94	&	3.80	&	3.55	\\
sat	&	110301.7+080253	&		&	1194	&	10	&	18.1	&	391	&	-13.13	&	12.23	&	3.09	&	2.85	\\
1554	&	111315.9+033926	&	0	&	6627	&	53	&	14.8	&		&	-20.17	&		&		&		\\
sat	&	111320.2+033713	&		&	6552	&	58	&	18.3	&	65	&	-16.64	&	11.62	&	1.40	&	0.94	\\
sat	&	111221.7+034540	&		&	6545	&	26	&	15.9	&	393	&	-19.06	&	12.49	&	2.27	&	1.81	\\
1613	&	113903.3-001222	&	5	&	5242	&	14	&	14.8	&		&	-19.59	&		&		&		\\
sat	&	113929.7+000700	&		&	5714	&	19	&	16.8	&	427	&	-17.77	&	14.05	&	4.05	&	3.53	\\
1614	&	113911.0+392002	&	3	&	7195	&	30	&	15.2	&		&	-19.83	&		&		&		\\
sat	&	113850.9+391716	&		&	7130	&	0	&	19.5	&	137	&	-15.51	&	11.83	&	1.74	&	1.21	\\
sat	&	113838.9+391556	&		&	7132	&	32	&	17.4	&	213	&	-17.61	&	11.99	&	1.89	&	1.36	\\
1655	&	115601.0-024315	&	0	&	5773	&	32	&	14.2	&		&	-20.40	&		&		&		\\
sat	&	115548.3-024434	&		&	5826	&	90	&	18.8	&	79	&	-15.82	&	11.40	&	1.08	&	1.05	\\
sat	&	115613.2-024029	&		&	5819	&	39	&	18.3	&	95	&	-16.31	&	11.37	&	1.05	&	1.02	\\
sat	&	115658.4-023804	&		&	5768	&	82	&	18.1	&	351	&	-16.47	&	10.31	&	-0.01	&	-0.05	\\
sat	&	115707.1-023159	&		&	5936	&	111	&	18.1	&	460	&	-16.54	&	13.15	&	2.83	&	2.80	\\
sat	&	115702.0-022853	&		&	5568	&	29	&	17	&	482	&	-17.50	&	13.39	&	3.07	&	3.03	\\
1657	&	115651.6+084252	&	3	&	10373	&	9	&	16.5	&		&	-19.37	&		&		&		\\
sat	&	115654.2+084643	&		&	10347	&	25	&	17.9	&	161	&	-17.95	&	11.14	&	1.23	&	-0.09	\\
1666	&	120236.5+410315	&	4	&	6154	&	10	&	13.82	&		&	-20.87	&		&		&		\\
sat	&	120219.9+410454	&		&	6190	&	3	&	16.9	&	87	&	-17.80	&	11.15	&	0.64	&	-0.47	\\
sat	&	120227.4+404913	&		&	6270	&	29	&	18.2	&	347	&	-16.53	&	12.76	&	2.25	&	1.14	\\
1672	&	120528.8+464647	&	4	&	9399	&	39	&	15.5	&		&	-20.13	&		&		&		\\
sat	&	120543.9+465548	&		&	9425	&	24	&	17.4	&	351	&	-18.23	&	11.45	&	1.23	&	-0.30	\\
1680	&	120932.9+170051	&	-2	&	6619	&	8	&	14.54	&		&	-20.40	&		&		&		\\
sat	&	120836.4+171243	&		&	6863	&	58	&	15.4	&	474	&	-19.67	&	13.52	&	3.20	&	2.94	\\
1704	&	121859.0-200451 	&	4	&	5398	&	34	&	14.31	&		&	-20.27	&		&		&		\\
sat	&	121857.5 -200029  	&		&	5765	&	210	&	18.2	&	94	&	-16.52	&	13.18	&	2.91	&	2.87	\\
1718	&	122435.0+015332	&	4	&	7031	&	53	&	15.9	&		&	-19.13	&		&		&		\\
sat	&	122502.8+014738	&		&	6753	&	33	&	18.1	&	255	&	-16.82	&	13.37	&	3.56	&	2.85	\\
1723	&	122545.5+514006	&	0	&	9190	&	32	&	14.9	&		&	-20.68	&		&		&		\\
sat	&	122541.2+514901	&		&	9490	&	26	&	18	&	328	&	-17.65	&	13.54	&	3.11	&	2.47	\\
1743	&	123344.9+521517	&	0	&	6695	&	36	&	14.54	&		&	-20.35	&		&		&		\\
sat	&	123418.0+520947	&		&	6788	&	29	&	18.4	&	200	&	-16.52	&	12.31	&	2.01	&	1.46	\\
1748	&	123600.1+541316	&	3	&	5450	&	11	&	14	&		&	-20.46	&		&		&		\\
sat	&	123538.1+541350	&		&	5511	&	51	&	17.7	&	71	&	-16.77	&	11.49	&	1.15	&	0.86	\\
1750	&	123713.5+492654	&	5	&	9061	&	51	&	14.5	&		&	-21.03	&		&		&		\\
sat	&	123755.2+493058	&		&	9090	&	29	&	17.9	&	285	&	-17.64	&	11.42	&	0.85	&	0.70	\\
1767	&	124426.2+370716	&	2	&	7007	&	6	&	13.9	&		&	-21.10	&		&		&		\\
sat	&	124313.5+370507	&		&	6557	&	50	&	17.7	&	409	&	-17.13	&	13.99	&	3.39	&	3.36	\\
1768	&	124428.1-030019	&	4	&	7003	&	11	&	14.9	&		&	-20.14	&		&		&		\\
sat	&	124504.7-030406	&		&	7309	&	28	&	16.8	&	276	&	-18.32	&	13.48	&	3.27	&	2.37	\\
1773	&	124611.5+484003	&	3	&	9325	&	34	&	15.3	&		&	-20.28	&		&		&		\\
sat	&	124500.3+483744	&		&	9371	&	42	&	16	&	445	&	-19.59	&	12.05	&	1.77	&	1.01	\\
1814	&	130743.7-123333	&	-2	&	6296	&	38	&	14.75	&		&	-20.13	&		&		&		\\
sat	&	130822.2-124053	&		&	6246	&	52	&	16.62	&	299	&	-18.23	&	11.95	&	1.73	&	0.66	\\
1820	&	130933.1+014023	&	4	&	5522	&	11	&	13.3	&		&	-21.25	&		&		&		\\
sat	&	130934.4+013725	&		&	5869	&	17	&	16.6	&	66	&	-18.08	&	12.97	&	2.31	&	2.52	\\
\end{tabular}
\end{table}

\begin{table}
\begin{tabular}{lccrrrrrrrr} \hline \\
\hline
1       &       2       &       3       &       4       &       5       &       6       &       7       &       8       &       9       &       10      &       11      \\
\hline
1828	&	131222.9-042011	&	1	&	3003	&	9	&	13.24	&		&	-19.99	&		&		&		\\
sat	&	131049.7-045101	&		&	2878	&	89	&	16.77	&	462	&	-16.36	&	12.90	&	2.75	&	3.08	\\
1838	&	131446.0+534913	&	1	&	4832	&	34	&	14.5	&		&	-19.71	&		&		&		\\
sat	&	131252.5+535722	&		&	4881	&	36	&	17.6	&	358	&	-16.63	&	12.02	&	1.98	&	1.79	\\
1846	&	131908.2+283025	&	3	&	6651	&	1	&	14.2	&		&	-20.68	&		&		&		\\
sat	&	131958.3+281449	&		&	6785	&	19	&	15.5	&	506	&	-19.42	&	13.03	&	2.60	&	2.34	\\
1855	&	132654.2+524513	&	-2	&	9036	&	39	&	14.4	&		&	-21.12	&		&		&		\\
sat	&	132630.3+524230	&		&	9484	&	37	&	17.9	&	163	&	-17.73	&	13.59	&	2.98	&	2.41	\\
sat	&	132611.8+524501	&		&	9145	&	45	&	17.8	&	231	&	-17.75	&	12.52	&	1.91	&	1.34	\\
sat	&	132807.5+525008	&		&	9184	&	47	&	16.6	&	436	&	-18.95	&	13.05	&	2.44	&	1.87	\\
1871	&	133439.2+040748	&	3	&	6776	&	33	&	14.7	&		&	-20.25	&		&		&		\\
sat	&	133451.0+041412	&		&	6765	&	6	&	17.6	&	190	&	-17.35	&	10.43	&	0.17	&	-0.29	\\
1873	&	133548.2+025956	&	2	&	6424	&	34	&	14.1	&		&	-20.73	&		&		&		\\
sat	&	133542.8+030007	&		&	6561	&	17	&	17	&	34	&	-17.87	&	11.88	&	1.43	&	1.47	\\
sat	&	133540.3+031735	&		&	6881	&	17	&	16	&	444	&	-18.98	&	14.04	&	3.59	&	3.62	\\
1915	&	135317.8+332927	&	2	&	2383	&	20	&	13.4	&		&	-19.26	&		&		&		\\
sat	&	134913.0+331900	&		&	2264	&	62	&	18.2	&	495	&	-14.33	&	12.89	&	3.02	&	2.88	\\
1919	&	135656.0+290952	&	2	&	2418	&	4	&	13.4	&		&	-19.29	&		&		&		\\
sat	&	135709.9+291310	&		&	2275	&	4	&	17.4	&	43	&	-15.13	&	12.02	&	2.14	&	1.88	\\
sat	&	135729.7+290332	&		&	2281	&	2	&	19.33	&	93	&	-13.20	&	12.31	&	2.43	&	2.17	\\
1922	&	135806.6-040509	&	5	&	7422	&	8	&	14.84	&		&	-20.37	&		&		&		\\
sat	&	135818.4-040141	&		&	7056	&	89	&	19.1	&	135	&	-15.99	&	13.33	&	3.02	&	2.64	\\
sat	&	135837.5-040135	&		&	7377	&	89	&	19.2	&	251	&	-15.99	&	11.78	&	1.47	&	1.09	\\
sat	&	135830.5-035253	&		&	7327	&	123	&	17.52	&	403	&	-17.65	&	12.64	&	2.33	&	1.95	\\
1987	&	142908.3+414951	&	1	&	5518	&	30	&	15	&		&	-19.44	&		&		&		\\
sat	&	142945.7+415158	&		&	5490	&	9	&	18.3	&	160	&	-16.12	&	11.20	&	1.26	&	0.78	\\
sat	&	142938.7+415747	&		&	5515	&	34	&	17.5	&	214	&	-16.94	&	9.36	&	-0.58	&	-1.06	\\
sat	&	142833.5+420224	&		&	5514	&	30	&	17.5	&	311	&	-16.93	&	9.77	&	-0.17	&	-0.65	\\
1997	&	143316.4+413901	&	3	&	5435	&	60	&	14.8	&		&	-19.62	&		&		&		\\
sat	&	143322.0+414023	&		&	5553	&	30	&	17.6	&	37	&	-16.86	&	11.79	&	1.78	&	0.85	\\
2012	&	143911.1+052149	&	0	&	1470	&	2	&	11.76	&		&	-19.92	&		&		&		\\
sat	&	143944.4+052112	&		&	1753	&	70	&	17	&	49	&	-15.08	&	12.66	&	2.54	&	2.65	\\
sat	&	144027.5+053155	&		&	1526	&	30	&	18	&	126	&	-13.76	&	11.64	&	1.51	&	1.63	\\
sat	&	143822.5+043648	&		&	1659	&	4	&	17.9	&	273	&	-14.07	&	13.08	&	2.95	&	3.07	\\
sat	&	143522.8+051636	&		&	1599	&	5	&	17.5	&	338	&	-14.34	&	12.85	&	2.72	&	2.84	\\
sat	&	144258.0+045322	&		&	1602	&	6	&	13.6	&	371	&	-18.27	&	12.88	&	2.75	&	2.86	\\
sat	&	144302.8+044556	&		&	1672	&	7	&	14.9	&	398	&	-17.05	&	13.28	&	3.15	&	3.27	\\
sat	&	144329.2+043153	&		&	1710	&	17	&	16.7	&	477	&	-15.29	&	13.51	&	3.38	&	3.50	\\
2018	&	144331.3+492335	&	3	&	9184	&	33	&	15.3	&		&	-20.32	&		&		&		\\
sat	&	144334.0+492451	&		&	9398	&	67	&	18.5	&	49	&	-17.16	&	12.42	&	2.13	&	1.65	\\
sat	&	144419.4+492443	&		&	9233	&	57	&	17.6	&	290	&	-18.01	&	11.90	&	1.61	&	1.13	\\
2023	&	144441.1-042019	&	5	&	13091	&	64	&	16.36	&		&	-20.28	&		&		&		\\
sat	&	144419.6-042038	&		&	13007	&	89	&	19.51	&	279	&	-17.08	&	12.36	&	2.09	&	0.79	\\
2029	&	144826.7+345953	&	3	&	8916	&	11	&	14.5	&		&	-21.02	&		&		&		\\
sat	&	144854.9+345214	&		&	8865	&	69	&	18.3	&	341	&	-17.19	&	12.00	&	1.43	&	1.06	\\
sat	&	144825.9+350932	&		&	8915	&	23	&	18.2	&	343	&	-17.29	&	9.21	&	-1.36	&	-1.73	\\
2034	&	145138.9+403557	&	3	&	5017	&	10	&	13.5	&		&	-20.76	&		&		&		\\
sat	&	145055.5+403126	&		&	5000	&	71	&	18.3	&	188	&	-15.95	&	10.75	&	0.29	&	-0.28	\\
sat	&	145001.6+402142	&		&	4929	&	38	&	18.3	&	467	&	-15.90	&	12.61	&	2.14	&	1.58	\\
2065	&	150654.8+001111	&	2	&	10506	&	28	&	15	&		&	-21.05	&		&		&		\\
sat	&	150654.5+001059	&		&	10166	&	21	&	18.6	&	9	&	-17.38	&	12.09	&	1.51	&	0.09	\\
sat	&	150702.0+001323	&		&	10596	&	21	&	18.5	&	119	&	-17.57	&	12.05	&	1.47	&	0.05	\\
2130	&	153626.3-665135	&	4	&	3294	&	14	&	12.4	&		&	-21.53	&		&		&		\\
sat	&	153621.4-665257	&		&	3228	&	113	&	16.4	&	19	&	-17.47	&	12.25	&	1.48	&	2.35	\\
2133	&	153722.9+203259	&	0	&	4604	&	48	&	15.1	&		&	-19.12	&		&		&		\\
sat	&	153741.3+204713	&		&	4545	&	5	&	18.2	&	273	&	-16.00	&	12.08	&	2.27	&	1.61	\\
\end{tabular}
\end{table}

\begin{table}
\begin{tabular}{lccrrrrrrrr} \hline \\
\hline
1       &       2       &       3       &       4       &       5       &       6       &       7       &       8       &       9       &       10      &       11      \\
\hline
2174	&	155601.6+302503	&	3	&	14125	&	47	&	15.1	&		&	-21.52	&		&		&		\\
sat	&	155622.7+302715	&		&	14241	&	25	&	18.2	&	317	&	-18.42	&	12.70	&	1.94	&	1.22	\\
2333	&	170128.2+634128	&	5	&	4974	&	19	&	14.6	&		&	-19.67	&		&		&		\\
sat	&	170026.3+634341	&		&	5150	&	20	&	16.6	&	143	&	-17.75	&	12.72	&	2.69	&	2.49	\\
sat	&	170457.2+634848	&		&	5090	&	17	&	16.5	&	481	&	-17.85	&	12.87	&	2.84	&	2.64	\\
2394	&	173044.8+562107	&	2	&	9361	&	30	&	15	&		&	-20.69	&		&		&		\\
sat	&	173029.8+561050	&		&	8961	&	22	&	17.8	&	391	&	-17.79	&	13.87	&	3.43	&	2.98	\\
2482	&	181649.3-571353	&	1	&	5000	&	20	&	13.8	&		&	-20.88	&		&		&		\\
sat	&	181637.2-571534	&		&	5009	&	46	&	17.4	&	47	&	-17.28	&	9.65	&	-0.86	&	-0.68	\\
2716	&	195647.6-602805	&	-2	&	3658	&	30	&	12.5	&		&	-21.21	&		&		&		\\
sat	&	195557.0-604356	&		&	3702	&	45	&	15.23	&	248	&	-18.52	&	11.79	&	1.15	&	0.88	\\
sat	&	195622.5-604533	&		&	3747	&	45	&	15.75	&	258	&	-18.03	&	12.40	&	1.76	&	1.49	\\
2777	&	202313.8-274252	&	-2	&	5808	&	27	&	12.95	&		&	-21.86	&		&		&		\\
sat	&	202306.5-273116	&		&	5920	&	45	&	16.8	&	271	&	-18.04	&	12.60	&	1.70	&	1.05	\\
2816	&	204052.1+003910	&	4	&	8256	&	16	&	15.8	&		&	-19.81	&		&		&		\\
sat	&	204119.8+003937	&		&	8211	&	41	&	17.4	&	228	&	-18.20	&	11.74	&	1.65	&	0.84	\\
2834	&	204952.2-070119	&	5	&	6225	&	13	&	15.4	&		&	-19.51	&		&		&		\\
sat	&	204929.4-064849	&		&	6480	&	19	&	18.1	&	340	&	-16.89	&	13.41	&	3.45	&	2.42	\\
2857	&	210203.5-061749	&	3	&	7941	&	16	&	14.4	&		&	-21.10	&		&		&		\\
sat	&	210159.0-062030	&		&	7892	&	22	&	18.6	&	93	&	-16.87	&	11.26	&	0.66	&	0.04	\\
2875	&	210826.9-054910	&	2	&	8551	&	24	&	14.8	&		&	-20.94	&		&		&		\\
sat	&	210814.0-055108	&		&	8613	&	24	&	17.4	&	128	&	-18.36	&	11.78	&	1.24	&	0.70	\\
2986	&	215656.7-252102	&	3	&	8847	&	64	&	14.73	&		&	-20.86	&		&		&		\\
sat	&	215659.1-252132	&		&	8727	&	89	&	18.03	&	26	&	-17.53	&	11.65	&	1.15	&	0.70	\\
sat	&	215726.7-252949	&		&	8936	&	64	&	17.35	&	391	&	-18.24	&	12.57	&	2.07	&	1.63	\\
sat	&	215608.7-252736	&		&	9146	&	64	&	18.07	&	447	&	-17.56	&	13.68	&	3.17	&	2.73	\\
sat	&	215635.4-250841	&		&	8662	&	45	&	16.77	&	467	&	-18.76	&	13.28	&	2.78	&	2.33	\\
2998	&	220344.5-274754	&	0	&	7115	&	24	&	14.61	&		&	-20.42	&		&		&		\\
sat	&	220338.5-275831	&		&	7063	&	64	&	18.1	&	303	&	-16.93	&	11.97	&	1.64	&	0.88	\\
3010	&	221048.6-435749	&	4	&	12903	&	45	&	15.68	&		&	-20.62	&		&		&		\\
sat	&	221140.4-435533	&		&	12945	&	89	&	19.41	&	498	&	-16.89	&	11.99	&	1.59	&	0.91	\\
3020	&	221606.2-302208	&	4	&	7734	&	11	&	14.58	&		&	-20.62	&		&		&		\\
sat	&	221555.3-301143	&		&	8010	&	89	&	19.35	&	329	&	-15.91	&	13.47	&	3.06	&	2.62	\\
3023	&	221820.5+133725	&	3	&	7733	&	30	&	15.6	&		&	-19.84	&		&		&		\\
sat	&	221842.0+132930	&		&	7792	&	49	&	18.8	&	292	&	-16.66	&	12.09	&	2.00	&	0.29	\\
3034	&	222342.4-263646	&	4	&	7612	&	31	&	13.96	&		&	-21.22	&		&		&		\\
sat	&	222408.3-263727	&		&	7518	&	64	&	19.1	&	177	&	-16.04	&	12.26	&	1.62	&	1.26	\\
sat	&	222421.7-264029	&		&	7548	&	123	&	18.49	&	289	&	-16.66	&	12.15	&	1.50	&	1.15	\\
sat	&	222253.2-263853	&		&	7548	&	89	&	18.84	&	342	&	-16.31	&	12.22	&	1.57	&	1.22	\\
3051	&	222955.4-081646	&	4	&	10758	&	31	&	14.64	&		&	-21.46	&		&		&		\\
sat	&	222947.1-081601	&		&	10872	&	24	&	18.3	&	94	&	-17.83	&	12.16	&	1.41	&	0.85	\\
sat	&	222959.2-081857	&		&	10563	&	24	&	18	&	103	&	-18.04	&	12.66	&	1.92	&	1.35	\\
3077	&	224142.1-324552	&	4	&	8483	&	45	&	15.87	&		&	-19.52	&		&		&		\\
sat	&	224100.5-324118	&		&	8502	&	89	&	18.07	&	334	&	-17.31	&	11.11	&	1.14	&	0.26	\\
3083	&	224424.4-000944	&	0	&	5070	&	13	&	13.6	&		&	-20.89	&		&		&		\\
sat	&	224446.8-000513	&		&	5344	&	70	&	18.4	&	146	&	-16.21	&	13.11	&	2.60	&	1.95	\\
3120	&	230632.4-252539	&	3	&	15841	&	45	&	16.3	&		&	-20.50	&		&		&		\\
sat	&	230628.9-252501	&		&	15879	&	64	&	18.5	&	64	&	-18.30	&	11.04	&	0.68	&	-0.25	\\
3194	&	234547.5-293104	&	0	&	10424	&	20	&	14.34	&		&	-21.52	&		&		&		\\
sat	&	234550.2-293041	&		&	10147	&	64	&	17.83	&	30	&	-17.97	&	12.43	&	1.67	&	0.88	\\
sat	&	234556.3-293224	&		&	10177	&	64	&	18.56	&	97	&	-17.24	&	12.84	&	2.08	&	1.29	\\
sat	&	234533.8-293254	&		&	9943	&	45	&	17.44	&	145	&	-18.31	&	13.60	&	2.83	&	2.05	\\
sat	&	234544.7-292431	&		&	10687	&	89	&	18.83	&	273	&	-17.08	&	13.35	&	2.58	&	1.80	\\
sat	&	234606.3-293818	&		&	10536	&	89	&	18.7	&	347	&	-17.17	&	12.72	&	1.95	&	1.17	\\
sat	&	234615.4-292512	&		&	10287	&	89	&	19.14	&	351	&	-16.69	&	12.90	&	2.13	&	1.34	\\
sat	&	234620.0-292307	&		&	10053	&	64	&	18.06	&	441	&	-17.72	&	13.86	&	3.09	&	2.30	\\
sat	&	234506.4-292444	&		&	10208	&	89	&	18.32	&	456	&	-17.49	&	13.40	&	2.64	&	1.85	\\
sat	&	234615.9-292057	&		&	10208	&	89	&	16.98	&	492	&	-18.83	&	13.44	&	2.67	&	1.88	\\
\end{tabular}
\end{table}}
\end{document}